\documentclass[aps,amsmath,twocolumn,amssymb,floatfixng,showpacs,
superscriptaddress,footinbib]{revtex4-1}

\usepackage{ulem}
\usepackage{graphicx,epstopdf}  
\usepackage{amsmath}
\usepackage{graphicx}
\usepackage{amsfonts}
\usepackage{color}
\usepackage{dcolumn}   
\usepackage{bm}        
\usepackage{amssymb}   
\usepackage{hyperref}
\usepackage{tikz}

\begin{document}

\title{Statistical prethermalization in randomly kicked  many-body classical rotor system}
\author{Aritra Kundu}\email{aritra.kundu@uni.lu}\affiliation{Department of Physics and Materials Science, University of Luxembourg, L-1511 Luxembourg}
\author{Tanay Nag}\email{tanay.nag@hyderabad.bits-pilani.ac.in}\affiliation{Department of Physics, BITS Pilani-Hyderabad Campus, Telangana, 500078, India}
\author{Atanu Rajak}\email{atanu@phy.iith.ac.in}\affiliation{Department of Physics, Indian Institute of Technology, Hyderabad 502284, India}

\newcommand{\Uf}{\vcenter{\hbox{ \begin{tikzpicture}
	\node[draw,rounded corners=5pt,minimum width=0.5cm,minimum height=0.5cm,fill=green!80,opacity=0.4,yshift=-1em] (circuitA1) at (0.5,0) {$ $};
	\end{tikzpicture}}}}
\newcommand{\Ui}{\vcenter{\hbox{ \begin{tikzpicture}
	\node[draw,rounded corners=5pt,minimum width=0.5cm,minimum height=0.5cm,fill=blue!20,opacity=0.8,yshift=-1em] (circuitA1) at (0.5,0) {$ $};
	\end{tikzpicture}}}}

 \newcommand{\Uni}{\vcenter{\hbox{ \begin{tikzpicture}
	\node[draw,rounded corners=5pt,minimum width=0.5cm,minimum height=0.5cm,fill=black!40,opacity=0.8,yshift=-1em] (circuitA1) at (0.5,0) {$ $};
	\end{tikzpicture}}}}

\date{\today}

\def\nn{\nonumber}
\def\redw#1{{\color{red} #1}}
\def\greenw#1{{\color{dgreen} #1}}
\def\bluew#1{{\color{blue} #1}}
\def\la{\langle}
\def\ra{\rangle}
\def\avg#1{\la #1 \ra}
\def\avr#1{\overline {#1}}
\newcommand{\std}[1] {\langle #1 ; #1 \rangle}

\def\ql{ {Q_{ l}} }
\def\qr{ {Q_{ r}} }

\newcommand{\cg}{C_\gamma}
\newcommand{\eg}{E_\gamma}
\newcommand{\ep}{\epsilon}
\newcommand{\g}{\gamma}
\newcommand{\be}{\beta}
\newcommand{\where}{\text{where}}
\newcommand{\for}{\text{for}}
\newcommand{\f}{\frac}
\newcommand\bea{\begin{eqnarray}}
\newcommand\eea{\end{eqnarray}}
\newcommand\beq{\begin{equation}}
\newcommand\eeq{\end{equation}}
\newcommand\p{\partial}
\newcommand\ie{{\emph{i.e.}}}
\newcommand{\bE}{\cx{E}}
\newcommand{\balpha}{\cx{\alpha}}
\newcommand{\angstrom}{\text{\normalfont\AA}}
\newcommand{\braket}[3]{\bra{#1}\;#2\;\ket{#3}}
\newcommand{\projop}[2]{ \ket{#1}\bra{#2}}
\newcommand{\ket}[1]{ |\;#1\;\rangle}
\newcommand{\bra}[1]{ \langle\;#1\;|}
\newcommand{\iprod}[2]{ \langle#1|#2\rangle}
\newcommand{\intl}[2]{\int\limits_{#1}^{#2}}
\newcommand{\logt}[1]{\log_2\left(#1\right)}

\newcommand{\mc}[1]{\mathcal{#1}}
\newcommand{\mf}[1]{\mathbf{#1}}
\newcommand{\mb}[1]{\mathbb{#1}}
\newcommand{\cx}[1]{\tilde{#1}}
\newcommand{\dx}[1]{\hat{#1}}
\newcommand{\blang}{\big \langle}
\newcommand{\brang}{\big \rangle}

\newcommand{\cmnt}[2]{\textbf{\#\#}{\color{#1}#2}\textbf{\#\#}}

\newcommand{\flap}{\mb{L}_{\bar{\kappa}}}
\newcommand{\flapfinitev}{\mb{L}^v}
\newcommand{\flapfinitem}{\mb{L}^p}
\newcommand{\fcurr}{\mb{A}}
\newcommand{\flapFull}{|\Delta|^{3/4}}
\newcommand{\tdir}{f}
\newcommand{\mzeta}{\chi}
\newcommand{\eline}{-----------------------------------------------------------------------------------\\}
\newcommand{\eqa}[1]{\begin{align}#1\end{align}}
\newcommand{\iu}{{i\mkern1mu}}

\begin{abstract}

We explore the phenomena of prethermalization in a many-body classical system of rotors under aperiodic drives characterised by waiting time distribution (WTD), where the waiting time is defined as the time between two consecutive kicks. We consider here two types of aperiodic drives: random and  quasi-periodic. We observe a short-lived pseudo-thermal regime with algebraic suppression of heating for the random drive where WTD has an infinite tail, as observed for Poisson and binomial kick sequences. On the other hand, quasi-periodic drive characterised by a WTD with a sharp cut-off, observed for Thue-Morse sequence of kick, leads to  prethermal region where heating is exponentially suppressed.  The kinetic energy growth is analyzed using an average surprise associated with WTD quantifying the randomness of drive. In all of the aperiodic drives we obtain the chaotic heating regime for late time, however, the diffusion constant gets renormalized by the average surprise of WTD in comparison to the periodic case.

\end{abstract}

\maketitle
\section{Introduction}
Periodically driven isolated many-body systems have recently attracted much interest due to their exciting outcomes compared to the corresponding equilibrium counterparts. Such time-periodic drives can generate new types of non-equilibrium orders like time-crystalline 
structures~\cite{else16floquet,khemani16phase,zhang17observation,yao17discrete} and various topological
phases~\cite{oka09photovoltaic,kitagawa11transport,lindner11floquet,thakurathi2013floquet,rudner13anomalous,vega19,seshadri19,nag14,nag17,nag19,nag21,Ghosh21,nag2021anomalous,Ghosh22}, generically known as  Floquet engineering~\cite{oka2019floquet,rudner2020band,weitenberg2021tailoring}. However, these systems absorb energy from the external periodic drives leading to the non-conservation of energy that results in a featureless infinite temperature state at late times. Consequently, 
in the intermediate time window, there can be 
interesting non-equilibrium steady states before these systems encounter a heat death hindering various novel applications.

The heating in periodically driven systems can be avoided by introducing 
strong disorder in the presence of interactions, thus creating many-body localized phases
~\cite{d13many,d14long,ponte15periodically,ponte15,lazarides15fate}. It also has been generically 
observed that the periodically driven systems, even in the clean limit show prethermalization behavior, i.e., 
the system initially equilibrates to a long-lived prethermal state at high frequency, followed by an exponentially 
small heating to the infinite temperature~\cite{choudhury14stability,bukov15prethermal,citro15dynamical,Mori15,chandran16interaction,canovi16stroboscopic,mori2016rigorous,lellouch17parametric,weidinger17floquet,abanin2017effective,else17prethermal,zeng17prethermal,abanin2017rigorous,Peronaci18,rajak2018stability,Mori18,Howell19,rajak2019characterizations,sadia2022prethermalization,Kundu21}. 
Recently, using NMR techniques~\cite{peng2021floquet} and ultracold atoms in driven optical lattice~\cite{rubio20}, 
the long-lived prethermal state has been observed in experiments.
The prethermal state is characterized by an effective time-independent Hamiltonian, 
that can be obtained from high-frequency Magnus expansion with time-period of the drive as a small parameter~\cite{mori2016rigorous,abanin2017rigorous}. 
Although the phenomenon of Floquet prethermalization is initially observed in non-integrable quantum systems with 
bounded operators where the heating bound is calculated following a rigorous approach,
the recent results show that these states can appear even for the classical counterparts~\cite{rajak2018stability,Mori18,Howell19,rajak2019characterizations,sadia2022prethermalization}. 
Quite remarkably, the classical prethermalization 
can even occur in systems with unbounded system variables and its origin is statistical in nature~\cite{rajak2018stability,rajak2019characterizations,sadia2022prethermalization}.

An important question then arises whether the Floquet prethermalization survives if one deviates from the perfect periodic limit. For quantum many-body systems, the recent reports suggest that the prethermal behavior can still be found for quasiperiodic ~\cite{dumitrescu2018logarithmically,zhao2021random} and some structured 
random drive protocols~\cite{zhao2021random}. The rigorous results on the bounds 
of the heating rates are obtained for continuous quasiperiodic driving~\cite{else2020long}, Thue-Morse quasiperiodic driving~\cite{mori2021rigorous}, 
as well as for the random multipolar driving~\cite{zhao2021random}.  Therefore, aperiodicity could also lead to Floquet prethermalization in quantum many-body systems. 
This motivates us to study whether a prethermal regime survives for classical interacting systems like quantum cases in the absence of perfect periodic drives. Such aperiodicities, yielding randomization of dynamics,  are also found to be useful for generating efficient quantum algorithms \cite{childs2019faster}.

Here, we consider three different protocols to observe the effect of aperiodicity on an interacting kicked rotor system. To study the system's dynamics beyond Floquet driving, we use the concept of waiting time ($t_w$) that determines the time between two consecutive kicks. In our analysis, the waiting time is not a constant number, unlike in the Floquet case, but is a random variable that follows a probability distribution. However, $t_w$ is always an integer multiple of Floquet time period. 
These protocols can be divided into two categories: for case one, the waiting time distribution (WTD) is unbounded, and for  case two, the distribution is bounded where there exists a finite extent of the waiting time. We examine the kinetic energy growth by varying the drive parameters.

In case one, we consider two purely random driving sequences, where protocol I follows the kicking sequence chosen from Poisson's WTD. In contrast, protocol II assumes a binomial kicking sequence. Both protocols I and II provide exact forms of WTDs with tails up to infinite value. The prethermal region marginally exists in this case while predominant heating is clearly noticed as kinetic energy always grows with time. We refer to such an arrest of kinetic energy growth as pseudo-prethermal regime where heating is algebraically suppressed.
For the case two, we consider the Thue-Morse sequence for the kicks as a quasi-periodic protocol III. This leads to a box distribution of the waiting time with equal probabilities without a tail. 
Interestingly, contrary to case one, we find a substantial regular prethermal region with exponential suppression of heating before the unbounded growth of the kinetic energy in case two.
By collapsing data of kinetic energy as a function of time for different values of system parameters, we find the expression of the diffusion constant and the heating time required to escape from the prethermal to chaotic state for all the cases of WTD. 
We characterize the randomness of drive with average surprise calculated from the entropy of WTD. Given a reference surprisal for all the protocols, we compare the universality of the results.  It has been shown that lifetime of the prethermal state reduces as the average surprise increases. We provide a theoretical argument using energy-time-like uncertainty relations to support our numerical findings.

\begin{figure}[h]
	\centering
\includegraphics[width=0.7\linewidth
]{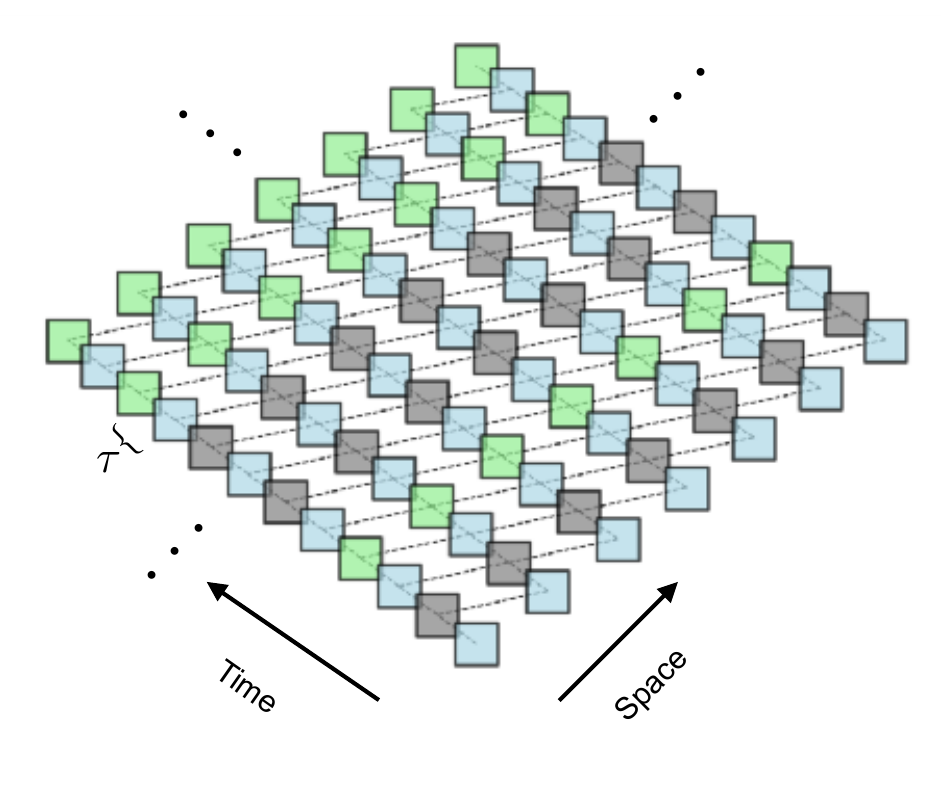}	
	\caption{Schematic plot for the equivalent circuit model for a Trotterized schedule for global random kicks. The particles undergo free evolution denoted by the green nodes and interact with the light blue or grey nodes. The grey nodes represent discrete times with missing kicks. Each free evolution is connected in space with its nearest neighbour nodes. The minimum time gap between the kicks is $\tau$ and the frequency of the missing kicks is characterized by a single parameter waiting time distribution (WTD). For the generalized Chirkov map model considered here, the equation of motion representation of the nodes is given in Eq. \eqref{eq_motion}. 
 } 
	\label{fig:fig1}
\end{figure}

\section{Model}
\label{model}
We consider here a classical many-body system of $N$ coupled oscillators with angle ($\theta_j$) and momentum ($\L_j$), associated with $j$-th oscillator, which evolves with a Trotterized schedule that can be either random or quasi-periodic. The discrete maps of $\L_j$ and $\theta_j$ between $n$-th and $(n+1)$-th events are given by
\eqa{\Ui:\L_j(n+1) &= \L_j(n)-\kappa \left( \sin(\theta_j-\theta_{j-1})-\sin(\theta_{j+1}-\theta_j)\right),\nonumber~\\ \Uni:\L_j(n+1) &= \L_j(n),\nonumber \\ \Uf:\theta_j(n+1) &= \theta_j(n) +  \L_j(n+1) \tau,
\label{eq_motion}}
where $\tau$ is the time between two consecutive events. The angle $\theta_j$ follows periodic boundary conditions $\theta_{j+N}=\theta_j$. These equations (\ref{eq_motion}) 
represent the generalization of the Chirikov standard map, a paradigmatic model for studying classical chaos. 
Considering the stroboscopic time $t$, which is defined just before the $n$-th event, i.e., $t=n-0^+$, $\kappa(n)$ is given as $\kappa(n) = \kappa G(n-0^+) $, where $G$ is a stochastic indicator function having the values $0$ or $1$ that determines whether the nearest-neighbor rotors interact or not at the $n$-{th} step. For a zero value of the indicator function, the system undergoes a free evolution with the momentum unchanged, while the position is shifted linearly in time by $\L_j(n) \tau$. In contrast, for the unit value of $G$, the momentum of a rotor changes in the next event by the interaction term, and the new value of the momentum shifts the angle. 

In an alternative way, the time-evolution between two consecutive events for the $j$-th rotor is given by a free evolution $U_j^f$ (green node) followed by either interacting evolution $U_j^i$ (blue node)  or non-interacting evolution $U_j^{ni}$ (grey node) .
Then the evolution of $j$-th rotor over the time duration $\tau$ is determined either by $U_j^+=U_j^iU_j^f$ or $U_j^-=U_j^{ni}U_j^f$ in a random fashion. We can define here the notion of waiting time $\tau_w$ that describes the time gap between two nearest instantaneous interactions between the nearest-neighbour rotors. In this work, we consider waiting time as a random variable that follows either a distribution or a quasi-periodic sequence. The minimum value of the waiting time is given by $\tau_w=\tau$, which describes two consecutive interaction events, and the corresponding dynamics are governed by the map $U_j(n,n+1)=U_j^+(n+1)U_j^+(n)$. Similarly $\tau_w=2\tau$ is defined by the map $U_j(n,n+1,n+2)=U_j^+(n+2)U_j^-(n+1)U_j^+(n)$ and thus goes on for any $\tau_w=l\tau$, $l$ being an integer.
As an example, the evolution of the total system up to time $t=n\tau$ is given by the map $U=\sum_jU_j^+(n)U_j^-(n-1)U_j^-(n-2)\cdots U_j^-(2)U_j^-(1)U_j^+(0)$. This mathematical description of the evolution of the total system can be represented by a schematic plot similar to a circuit model, as shown in Fig.~\ref{fig:fig1}.

The above equations of motion in Eq.~(\ref{eq_motion}) lead  to a Hamiltonian of the form
\beq
H = \sum_{j=1}^N \Big[\frac{\L_j^2}{2} - \kappa \Delta(t) \cos(\theta_{j+1} - \theta_j)\Big],
\label{mkr_eq}
\eeq
where the kick strength or alternatively the interaction strength $\kappa$ is a constant, but the delta function kicks to take the form, $\Delta(t)=\sum_n\delta(t-\sigma_n)$ with $\sigma_n = \sum_i^n \tau_w^i$ being the sum of intermediate 
waiting time random variable $\tau_w^i$, associated with $i$-th missing kick up to $n$-th event. Importantly, $\tau^i_w$ is chosen with probability $P_{\lambda}(\tau^i_w)$, corresponding to the parameter $\lambda$, from the WTD.   A random Trotterized driving sequence is thus  characterized by WTD assuming a lower cut-off for the waiting time, i.e., ${\rm min}(\tau^i_w)=\tau$ and, in general, $\tau^i_w=l\tau$ with $l$ being a positive random integer with an underlying probability distribution. It is safe to assume that $\sigma_0 = 0$.

Another alternative viewpoint of the waiting times is time-of-arrival, and probability densities are similar to clock operators \cite{hegerfeldt2010}. The WTD can be computed by plotting the frequency of missing kicks in the limit of a large number of epoch.
It can be shown that the system in Eq.~\eqref{eq_motion} has only one dimensionless 
parameter $K=\kappa\tau$, that determines the dynamics of the system. We consider $\tau=1$ throughout this paper without any loss of generality.
For the delta function WTD, the problem reduces to the Floquet limit of a standard many-body kicked rotor model that describes the transition from regular dynamics to classical chaos~\cite{kaneko89diffusion,konishi90diffusion,falcioni91ergodic,chirikov1993theory,chirikov97arnold,mulansky11strong,rajak2020stability},
and was shown to support a prethermal regime whose lifetime 
is exponentially large for high frequency and small amplitude of the drive.

In this work, we explore the effect of random waiting time on the dynamics of the many-body kicked rotor or generalized standard map.
To examine both the prethermal and the chaotic behaviour of the system, we numerically compute time dynamics of the average kinetic energy per rotor, 
$E_{\rm kin}(t)
=(1/N)\sum_{j=1}^N\avr{\langle \L_j^2(t)/2\rangle}$ using Eq.~(\ref{eq_motion}). The symbol $\langle..\rangle$ indicates the averaging over the initial conditions where $\L_j(t=0)=0$ for all $j$, and $\theta_j(t=0)$ are chosen from a uniform distribution between $0$ and $2\pi$. In addition, the overhead bar denotes the averaging over different configurations of the waiting times.

\section{Waiting time distributions (WTD) and driving}
\label{RAD}
We are interested in finding the effect of aperiodic or random kicks in the dynamics of the many-body kicked rotor, given that the perfect periodic kicking results in a long-lived prethermal state. In this context, we consider three aperiodic driving protocols and discuss their imprints on the subsequent dynamics.  We note that WTD may or may not have a sharp cut-off and depending upon this nature one can categorize the driving protocol. To be precise,  the first two driving protocols belong to the case one scenario where the tail of WTD extends to infinity having no sharp cut-off. On the other hand,  the last protocol corresponds to the case one situation where WTD shows a box profile with a sharp cut-off. Below we demonstrate the driving protocols in details. 


\subsection{Protocol I: Poisson clock/Exponential distribution}
\label{poisson_dist}


We here demonstrate the first driving scheme namely protocol I to analyze case one.  
In this case, the waiting time between two kicks follows a discrete exponential distribution and from there we can draw a typical kicking sequence. The normalized WTD for this case takes the form $P_{\gamma}(l)=(e^{\gamma}-1)e^{-\gamma l}$, where $l=\tau_w/\tau$ is an integer and $\gamma$ is a real parameter that determines the spreading of the distribution. The average and variance of the integer waiting time can be calculated as $\avg{l}=e^{\gamma}/(e^{\gamma}-1)$ and $\std{l} =\text{cosech}(\gamma/2)/2$, respectively. The average suprisal of WTD is defined as $S_\gamma=-\sum_l P_\gamma(l)\ln P_\gamma(l)$  determines the randomness of the distribution and is given as $S_\gamma= -\frac{\gamma  e^{\gamma } }{1-e^{\gamma }}-\log \left(e^{\gamma }-1\right)$. 
The histogram of waiting time with corresponding exponential distribution for $\gamma=1$ in Fig.~\ref{fig:poisson_ke}(a).  See Appendix \ref{appendix_A} for a detailed discussion.

The time-evolution of the kinetic energy for the exponential WTD is shown in Fig.~\ref{fig:poisson_ke}(b) for two values of $K$ with four different $\gamma$ for each $K$. The system shows an unbounded diffusion in kinetic energy,  preceded by a short-lived prethermal state when the kinetic energy is  almost constant.
 
The lifetime of the prethermal state depends both on $K$ and $\gamma$; as anticipated, the lifetime of the regime depends on the modified Fermi-golden rule for Markovian waiting time, it decreases with $K$ and the average surprise, determined by $\gamma$.  This can be explained using the form of the WTD. As $\gamma$ increases, the spreading of the distribution decreases and finally in the limit of $\gamma\rightarrow\infty$, the distribution converges to the delta function representing a nearly Floquet scenario.
Therefore, for a particular value of $K$, the system faces less temporal randomness as the value of $\gamma$ increases and the prethermal state's lifetime increases.
On the other hand, for $\gamma \to 0$, WTD gets flattened leading to a rapid rise of kinetic energy and thus disrupting the prethermal behavior substantially.

To get further insight, we make data collapse of the kinetic energy curves for different values of $K$ and $\gamma$ by rescaling the time axis with $S_\gamma K^2t$ (see Fig.~\ref{fig:poisson_ke}(c)). In the chaotic regime, the average kinetic energy shows diffusive behavior $E_{\rm kin}=D(K,\gamma)t$, where $D(K,\gamma)$ is the diffusion constant of the system. 
From our numerical results, we find that the diffusion constant of the system in the chaotic regime for this driving protocol can be obtained as $D(K,\gamma)\approx S_\gamma K^2$. Comparing the diffusion constant with the Floquet case, we can observe that, for the present scenario, the diffusion constant is renormalized by a factor $S_\gamma$. 
We now numerically determine the lifetime of the prethermal state $t^*$ of the system by analyzing the growth of kinetic energy as a function of the time $t$.

To measure the prethermal time quantitatively, we fit the kinetic energy $E_{\rm kin}$ up to time $t$ by a scaling function $t^{\alpha}$ with $\alpha<1$ for different values of $K$ and $\gamma$. We assume the time as $t^*$, designated by the upper limit of the fitting range, for which $\alpha\approx1$ given $K$ and $\gamma$ fixed. 
Having obtained the heating time $t^*$ for several values of $K$ and $\gamma$, we plot ${\rm log}(t^*)$ vs. ${\rm log} (1/K)$ in  Fig.~\ref{fig:poisson_ke}(d) keeping three representative values of $\gamma$. The lifetime of the prethermal state shows a scaling relation $t^*\sim1/K^a$, where the exponent $a$ varies between $1.9-2$ for three different values of $\gamma$.

Following a quantitative analysis, we observe that the prethermal lifetime of the system increases with $\gamma$ as expected. This indicates the growth of kinetic energy is algebraically suppressed with $K$ as compared to the exponential suppression in the regular prethermal region observed for periodically kicked case \cite{rajak2019characterizations}. We define this state as pseudo-prethermal region. This regime is also distinguished through the dynamics of average phase slips
defined as the average number of times the relative angle between adjacent rotors crosses $2\pi$:
$\avg{\phi} = \frac{1}{N} \sum_i {\rm Mod}[\phi_{i+1}-\phi_i,2\pi]$. It measures the dynamic transitions that occur in evolution as a proxy. The average phase slips changes from having a $t^{1/2}$ behaviour in the prethermal regime to a $t^{3/2}$ behaviour in the heating regime supporting distingusiblity of the prethermal and heating regimes by the exponents of phase slip dynamics, See. Appendix \ref{app:phaseslip}.

\begin{figure}
    \centering
    
  \includegraphics[scale=0.5]{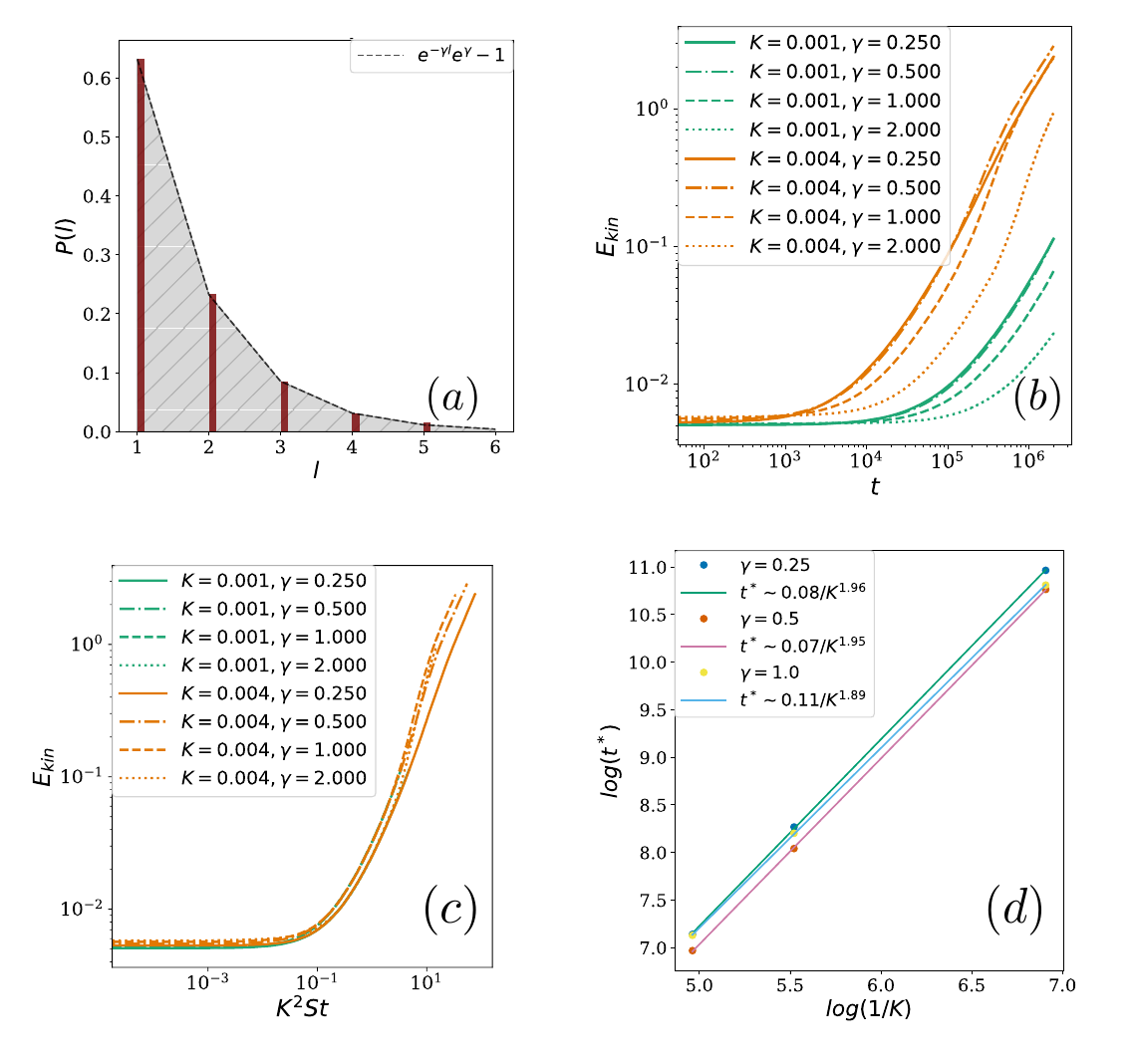}
    \caption{(a)  WTD for Poisson distribution with  $\gamma = 1$ having the surprisal $S_{\gamma}=S\approx1$. (b) Variation of the average kinetic energy per rotor 
    with time for different values of $K$ and $\gamma$ with the exponential WTD. 
    (c) Data collapse of the kinetic energy curves with rescaled time axis by $SK^2t$. (d) Scaling of the heating time $t^*$ with $1/K$ in log-log plot. 
    }
    \label{fig:poisson_ke}
\end{figure}

\subsection{Protocol II: Binomial kicking sequences}
\label{binomial}

To analyse case one, we demonstrate the second driving scheme, protocol II.  
In this instance, the kicking sequence is determined by a combination of kick and no-kick following a binomial probability distribution at each stroboscopic time. We can find the WTD from the kicking sequences. It can be noted that this is an alternative approach as compared to 
protocol I in Sec.~\ref{poisson_dist}, where the kicking sequence is determined from the WTD. However, for both cases, the waiting time can increase to an infinite value, although the probability for such cases is negligibly small. The randomness of the drive is quantified by a probability $p$, which determines the occurrence of 
kick, whereas $(1-p)$ is the probability of having no kick. For $p=1$, the system is periodically kicked and it has a long-lived prethermal state. The other extreme limit $p=0$ corresponds to the free evolution of the system, and the kinetic energy is always a conserved quantity. The maximum randomness in the system dynamics is induced for $p_c=1/2$, where the binary entropy function becomes maximum.

For this driving protocol, the analytic form of the WTD is found as follows: The probability of $l$ consecutive missing kicks is given by $(1-p)^l$, where $(1-p)$ is the probability of one missing kick. This leads to the distribution of waiting time having the form {$P(l)=C(1-p)^{l}$}, where $l=\tau_w/\tau$ is an integer and $C=\frac{p}{1-p}$ is the normalization constant.
The average and the variance of integer waiting time for this setting are given by $\langle l\rangle=\frac{1}{p}$ and $\std{l}=\frac{1-p}{p^2}$, respectively. In this case, the surprise can be calculated as $S_p=-\frac{1}{p} \left(\ln[p^p(1-p)^{(1-p)}]\right)$. As mentioned before, to plot the WTD, we consider here $p=0.63$ so that $S_p\approx1$ (see Fig.~\ref{fig:binomial_ke}(a)). See Appendix \ref{appendix_A} for a detailed discussion.

The kinetic energy evolution for the kicking sequence mentioned above is shown in Fig.~\ref{fig:binomial_ke}(b) for different values of $K$ and $p$. Although the model in Eq.~(\ref{mkr_eq}) shows eventual unbounded chaotic diffusion, it has a short-lived prethermal state, to be precise, pseudo-prethermal state,  for very small values of $K$ when the kinetic energy is almost constant; the lifetime of the state depends both on $K$ and $p$. 
We observe a subtle difference: for $p<p_c$, there is an initial rise in the kinetic energy before entering into the {pseudo-prethermal} regime, which is absent in the kinetic energy curves for $p>p_c$. 
To study the time-evolution of the kinetic energy, we consider here two values of $K$ and for each $K$, three different values of $p$ are assumed. We observe that for a particular value of $K$, the system starts heating up first for $p=1/2$, i.e., when the randomness in the kicking sequence is maximum. Otherwise, the lifetime of the pseudo-prethermal state increases with an increasing value of $p$. The kinetic energy curves for a single $K$ with different $p$ but symmetric about $p=0.5$ shows a perfect collapse at large $t$, i.e., when the system behaves chaotically. It indicates that the chaotic dynamics of the system is self-similar about $p=0.5$.

We rescale the time axis to $K^2S_{|p-1/2|}t$ while examining the kinetic energy curves 
for different $K$ and $p$, except for $p=0.5$. We find that the curves show a good collapse in the chaotic regime compared to the pseudo-prethermal region (see Fig.~\ref{fig:binomial_ke}(c)).  
We observe that the energy curves for $p=0.5$ and any $K$ can not be merged with the same rescaling of the time axis. In the chaotic regime, the average kinetic energy shows diffusive behavior $E_K=D(K,p)t$, where $D(K,p)$ is the diffusion constant of the system.
Consequently, the diffusion constant for this driving protocol is given by $D(K,p)\approx K^2S_{|p-1/2|}$. One can understand $K^2$ dependence of $D(K,p)$ using the approximation of independent rotors in the heating regime. Due to the aperiodic drive using binomial sequence, the diffusion constant is modified by the parameter $p$ that determines the intensity of aperiodicity.  
From the expression of $D(K,p)$, we observe that the diffusion constant is symmetric about the critical probability $p_c=1/2$.

We now numerically determine the prethermal lifetime $t^*$ of the system by plotting the kinetic energy curves as a function of the time $t$. Using the method described in Sec.~\ref{poisson_dist}, we numerically determine the lifetime $t^*$ for different values of $K$ and $p$. The prethermal lifetime $t^*$ as a function of $1/K$ in log-log scale is shown in Fig.~\ref{fig:binomial_ke}(d) for three different values of $p$. The straight line fit for the data points
suggests a power-law function of the heating time as $t^*\sim1/K^a$. For all the values of $p$, the scaling exponent $a$ is $2$ within numerical precision. Among these three values of $p$, the lifetime of the prethermal state is lowest for $p=0.5$, i.e., when the temporal randomness is maximum and the system gets heated up more quickly than the other two values of $p$. On the other hand, the lifetime of the prethermal state is larger for $p=0.2$ compared to the case of $p=0.8$, i.e.; heating takes place early for $p=0.8$ compared to  $p=0.2$ leading to a further reduction of the pseudo-thermal region. This can be explained using the equations of motion in 
Eq.~(\ref{eq_motion}). For $p=0.2$, the process involves a larger number of missing kicks compared to the case with $p=0.8$; therefore, to reach the resonance condition, the first case takes more time than the second one since the momentum remains unchanged during the missing kicks (see Eq.~(\ref{eq_motion})).  

\begin{figure}
    \centering
   \includegraphics[scale=0.5]{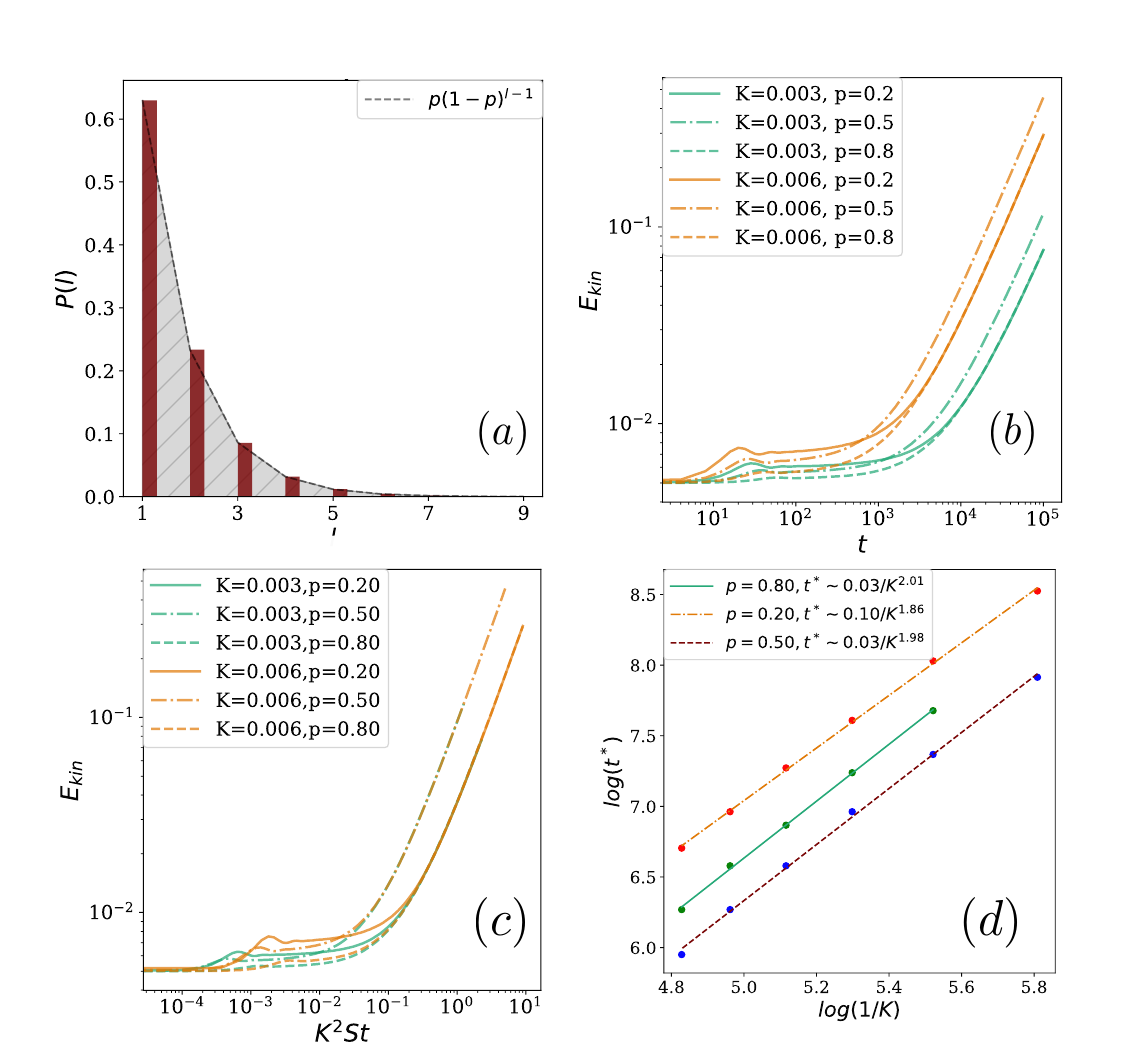}
    \caption{(a) The blue bars indicate the normalized frequency of the integer waiting time $l$ corresponding to kicks from a Binomial distribution with probability of kicks, $p=0.63$ and surprisal $S=1.046$. The corresponding normalized distribution perfectly matches the analytically found expression indicated with a dotted line. (b) Variation of the average kinetic energy with time $t$ for different values of $K$ and $p$ for the binomial kicking sequence. (c) The collapse of the kinetic energy curves for different $K$ and $p$, except $p=0.5$, by rescaling the time axis as $K^2St$. It shows an excellent collapse in the chaotic regime. (d) Heating time $t^*$ as a function of $K$ for binomial kicking protocol with different values of $p$. It scales as $t^* \sim 1/K^{\alpha}$ with  the driving frequency ($1/K$). Here $\alpha$ can be equated to $2$ within numerical precision. 
    } 
    \label{fig:binomial_ke}
\end{figure}

\subsection{Protocol III : Random Multipolar and Thue-Morse driving}
\label{thue-morse}

\begin{figure}
    \centering
      \includegraphics[scale=0.5]{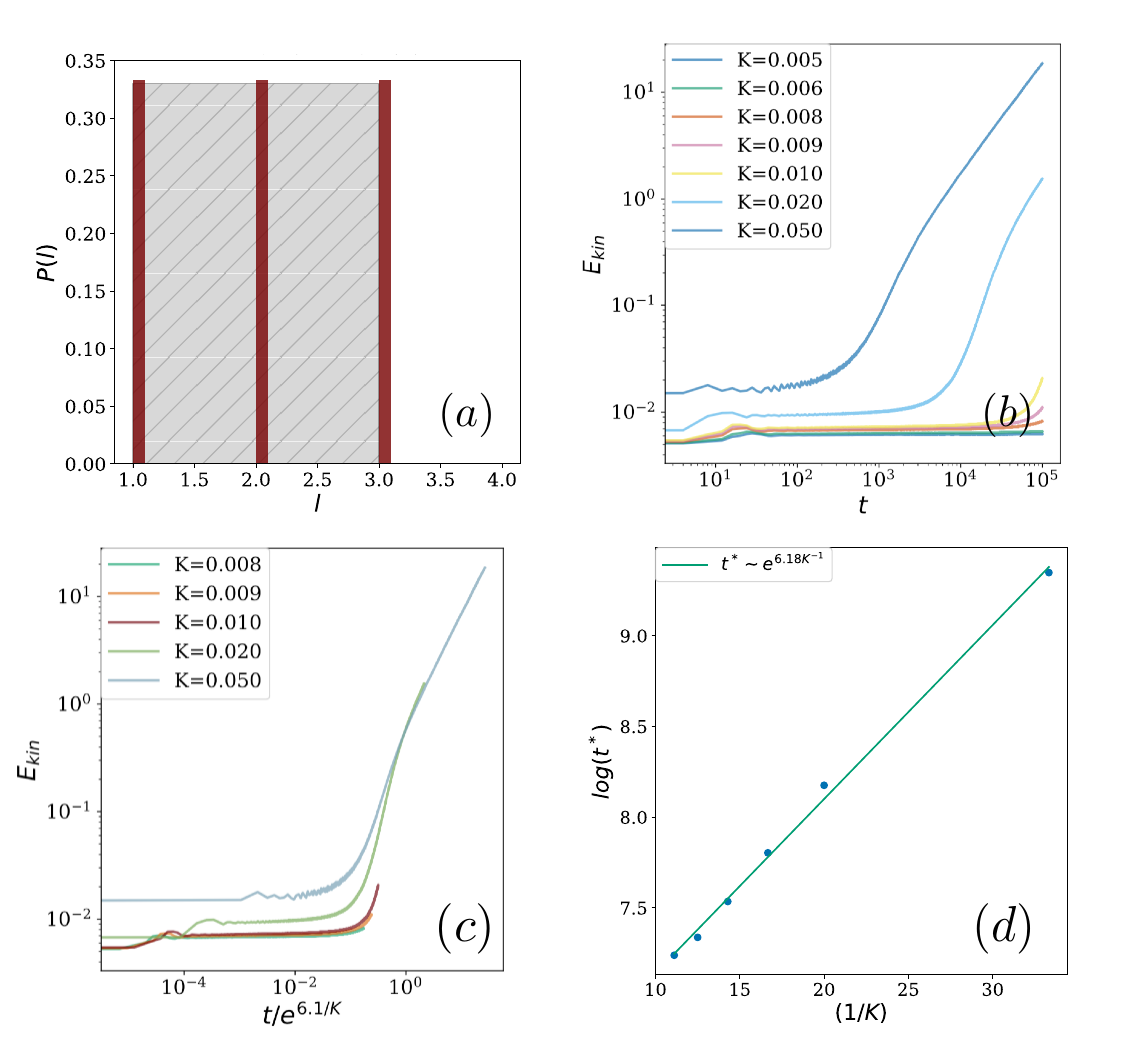}
    \caption{(a) The normalized WTD for this sequence shows three peaks at $l=1, 2$ and $3$ in contrast to the previous two cases in Figs.~\ref{fig:binomial_ke}, and \ref{fig:poisson_ke}, where the integer waiting time can be extended up to $\infty$ with small but non-zero probability. (b) The average kinetic energy per rotor for different values of $K$, when the system is driven according to the Thue-Morse driving schedule. (c) Data collapse of the kinetic energy curves in the chaotic regime for different values of $K$ with rescaled time axis by the $e^{c/K}$, where $c=6.18$.  (d) The heating time $t^*$ with $1/K$ in semi-log plot using the same scaling.
      }
    \label{fig:thue-morse_ke}
\end{figure}

We now examine case two, considering another type of driving sequence that provides a finite cut-off to the waiting time distribution, unlike the previous case one. In this context, we investigate the system dynamics driven by quasi-periodic Thue-Morse or structured binary random sequences defined by random multipolar drives. A multipolar driving is comprised of 
a structured sequence of multipolar blocks while inside a given block, there can be $m$ number of entries with $m$ being the order of the multipole.  These drives exhibit correlations between the blocks.
For $m=0$, the drive is monopolar and generated by a random choice with equal probability between two binary options $\{s_0^+,s_0^-\} = \{0,1\}$, where $0$ and $1$ signify the absence and the presence of kicks at any stroboscopic time, respectively. We note that the 
{monopolar} drive with $m=0$ exactly corresponds to the case of the binomial kicking sequence with $p=0.5$ as discussed in Sec.~\ref{binomial}. Similarly, the dipolar driving corresponds to the random choices between two elementary dipolar blocks $\{s_1^+,s_1^-\} = \{(0,1),(1,0)\}$. In continuation, for any general multipolar drive $m$, the sequence is recursively formed using two different $(m-1)$-th blocks, $\{s_m^+,s_m^-\} = \{(s_{m-1}^+,s_{m-1}^-),(s_{m-1}^-,s_{m-1}^+)\}$. For $m=0$, the WTD has only an infinite extent as in the case of binomial kicking, otherwise, the distribution is discrete with finite extent for any non-zero $m$.
The $m \to \infty$ limit corresponds to the quasi-periodic Thue-Morse (TM) kicking sequence, $011010011001011010010110\dots$ as obtained using 
$A \to A \bar{A} \to A \bar{A} \bar{A} A \to \cdots$ iterations. 
From this quasi-periodic sequence, we observe that the waiting time can take three values, $\tau_w=\tau$, $2\tau$ and $3\tau$ with equal probabilities. Therefore, the normalized integer WTD can be written as $P(l)=\sum_{i=1}^3\frac{1}{3}\delta(l-i)$ which is shown in Fig.~\ref{fig:thue-morse_ke}(a). The average and variance of the corresponding distribution are given by $\langle l \rangle = 2$, $\langle l;l \rangle = 2/3$, respectively. The surprise of the distribution can be calculated as $S=1.09861$ which is close to unity that we intended to keep for all the driving sequences to plot the WTD. See Appendix \ref{appendix_A} for a detailed discussion.

Recently, this sequence has been used to drive quantum systems in the context of finding long-lived 
prethermal states before eventual heating, even without perfect periodic drive protocols~\cite{mori2021rigorous}. 
For a generic quantum system, the heating rate is suppressed faster than a power law 
of driving frequency, unlike to the random driving case. A rigorous heating bound has also been 
observed for the generic quantum systems having operators with local norm bound. The existence of the prethermal behaviour for the system with the Hamiltonian in Eq.~(\ref{mkr_eq}) is established in Ref.\cite{yan2023prethermalization} for the aperiodic drives with random multipolar and quasi-periodic TM kicking sequences. In the above work, the authors have considered symmetric kicking strength to find the temperature and the lifetime of the prethermal state for TM quasi-periodic drive. We assume here asymmetric kicking strength for TM drive and wish to investigate the difference in prethermal behaviour of the system compared to the symmetric case as done in Ref.\cite{yan2023prethermalization}.

The time-evolution of the average kinetic energy for the asymmetric TM driving is shown in Fig.~\ref{fig:thue-morse_ke}(b) for different values of the kicking strength $K$. Unlike the previous two cases, the Thue-Morse driving does not have any parameter that determines the strength of temporal disorder in the system and therefore the randomness is structured in this case. This is reflected in the prethermal behavior of the system since it looks prethermal state is more robust for the present case compared to the previous two cases (see Figs.~\ref{fig:poisson_ke}(b), \ref{fig:binomial_ke}(b), and \ref{fig:thue-morse_ke}(b)). One can hence refer to this phase as the regular prethermal region opposing the 
pseudo-prethermal region found in case one.  To further analyze the prethermal behavior, we adopt an exponential rescaling of time $t \to t e^{-c/K}$ which shows nice data collapse for smaller values of $K$, see Fig. \ref{fig:thue-morse_ke}(c). 
The diffusion takes place in the chaotic region after experiencing the prethermal region where energy increases with time. In this region, the renormalized time $t e^{-c/K}$ does not lead to a data collapse for the chaotic regime, indicating a possible algebraic dependence on $K$.

 The lifetime of the prethermal state  $t^*$ (defined in Sec.~\ref{poisson_dist}) is plotted with $1/K$ in the semi-log scale in Fig.~\ref{fig:thue-morse_ke}(c) and found that it scales as $t^* \sim e^{c/K}$, where $c$ is a numerical constant. The present case indicates the exponential suppression of the heating for the regular prethermal region as in opposition to the algebraic suppression of heating within pseudo-prethermal region obtained for case one. This is similar to the periodic Flqouet prethermal phase where heating is exponentially suppressed. 
The expression of the time scale corresponding to the prethermal state for TM drive is different from Ref.~\cite{yan2023prethermalization}, where it takes the form $t^* \sim e^{\log(1/K)^2}$. However, the main difference between these two studies leads to two different prethermal time scales; the driving pattern between two binary kick strengths is either $0$ or $K$, i.e., a combination of kicks and missing kicks and, for them, two binary kick strengths are $\pm K$ having zero missing kicks.

\section{Aperiodic driving: theoretical arguments for numerical observation}
\label{stoc_thermo}

In the previous sections, we discuss our numerical results on the dynamics of the average kinetic energy of the rotor for random and quasi-periodic drives. For case one (II), we observe the existence of the pseudo- (regular) prethermal state and find the corresponding heating time 
that scale with the surprise of the WTDs. 
In this section, we discuss the case of a continuous drive. Then, using energy-time like uncertainty relations, we give a heuristical argument for scaling of prethermal timescales with the average surprise of the WTD.

\textit{Continuous random drive:} 
Although our work deals with discrete random drives before diving into those particular cases, let us first anticipate the case of continuous random drive when the coupling $\Delta(t)$, is a Gaussian white noise with unit average and variance $\gamma$. The evolution of an averaged observable $\avr{A}$ is given using quantum-classical correspondence \cite{Martinez2023PRL}[also see in Appendix \ref{app:doublePoisson}] as

\eqa{\avr{A}(t) \sim e^{i\mc L^\dagger t} \avr{A}(0) \label{eq:avgevo}}
where $i \mc{L}$ is the generator of the  
$i\mc L^\dagger [\circ]= \{H_0 +\gamma H_1,\circ\} -2\gamma \{H_1,\{H_1,\circ\}\}$
where $H_0 = \sum_{j=1}^N \L_j^2/{2} $ is the kinetic energy and $H_1 = \kappa \sum_j \cos(\theta_{j+1} - \theta_j)$ and $\{,\}$ denotes the classical Poisson brackets.
The total momentum ($\avr{\sum_j \L_j}$) is the constant of motion of the system. In the heating regime, the total kinetic energy grows diffusively  $\avr{\sum_j \L_j^2} \sim 2 \gamma \kappa^2 t$ and the diffusion constant from Fermi golden rule is proportional to the squared strength of the interaction.  
Now, let us turn to the case of the discrete process with $n$ interactions, one expects that this trend will hold, i.e. $\avr{\L_j^2(n)} \sim \alpha n$, where $\alpha$ is proportional to the variance of the WTD of the drive.  However, a better scaling for the diffusion constant of the heating regime from the numerical simulations is obtained when $\alpha$ is proportional to the average surprise of the WTD of discrete drive. Next, we give a heuristic argument for the same.

\textit{ Heuristic arguments on prethermal temperature for discrete random drive: } 
In this section, we give heuristic arguments to get the prethermal temperature for a discrete random drive. We use, a variant form of uncertainty relation of energy and time type \cite{nicholson2020time}, $\avg{\Delta l}\avg{\Delta H}/T^*\sim n\sim \tau_{eff}/\tau$ helps us to understand the observed scaling behaviour of the prethermal lifetime. Here $\Delta l = \sqrt{\avg{l;l}}$ is the standard deviation of the WTD, $\Delta H$ is the energy fluctuation in the prethermal state, and $n$ is number of events. The system escapes from the prethermal state for $n\sim \tau_{\rm eff}/\tau$ defined as the effective number of interactions it takes for the average energy to be equal to the variance of the energy. 
The time scale $\tau_{\rm eff}$ can be interpreted as the lifetime of the prethermal state. For randomly driven cases, the lifetime of the prethermal states can be defined with $\Delta l\sim O(1)$, $\Delta H \sim O(1/\tau^2)$. Using the above uncertainty relation, we find that the lifetime of the prethermal scale is proportional to the inverse of the prethermal temperature ($\tau_{eff} \sim t^*\sim 1/\tau T^*$). From numerical simulations, the lifetime as $ t^*\sim \tau/SK^2$. This implies
the prethermal temperature to be $T^* \sim SK^2/\tau^2$. This trend is also consistent with the numerical results presented in Fig.~\ref{fig:poisson_ke}(c) and Fig.~\ref{fig:binomial_ke}(c). 

\begin{figure}
    \centering
    \includegraphics[scale=0.52]{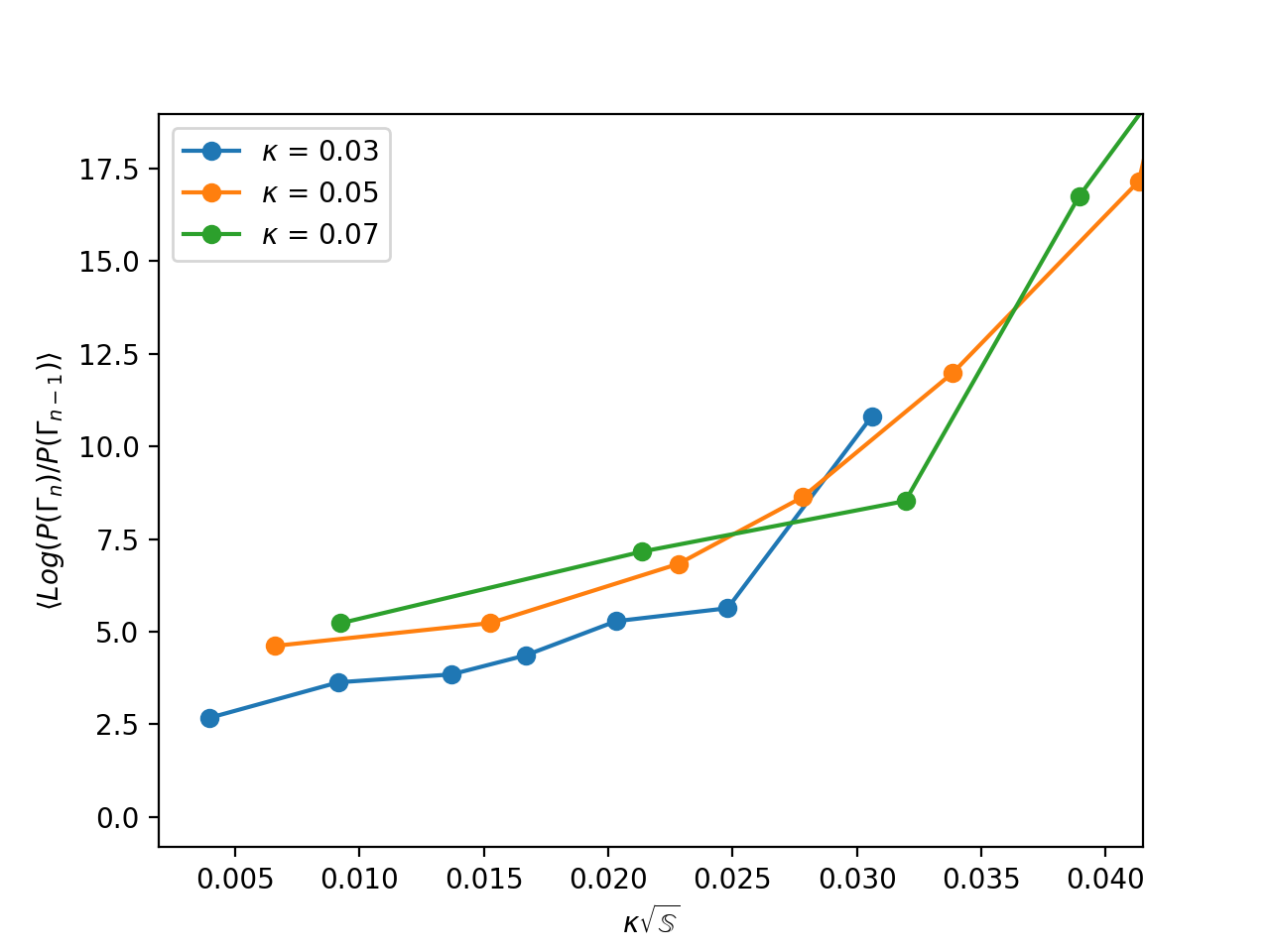}
    \caption{We show the change in the phase space entropy $\Delta \mb S_n$ with the scaled surprisal $K\sqrt{S}$.  }
    \label{fig:kick-strength-disorder}
\end{figure}

To investigate further, we consider the phase space at $n$th kick as $\Gamma(\{\L_n,\theta_n\}) \equiv \Gamma_n$ to analyze the connection between average phase space entropy, $\mb S_n = \int \mb P(\Gamma_n) \log(\mb P(\Gamma_n)) d\Gamma_n$) and the entropy supplied by the randomness of the kick through average surprise ($S$). We measure the change in phase space entropy under two consecutive events as a function of $S$ for Poisson WTD with different kick strengths as shown in Fig.~\ref{fig:kick-strength-disorder}. By rescaling the $x-$axis, we find in the high-frequency limit, the average change in entropy of the phase space ($\Delta \mb S_n$) is proportional to the $K\sqrt{S}$. Combining the above observations, we find that the lifetime of the prethermal state decreases as $t^* \sim \tau/\Delta \mb S_n ^2$.

\section{Discussion}

In this study, we have considered a circuit model where the particles go through free evolution followed by random or quasi-periodic kicks. In the process, the kinetic energy of the particles either remains constant or changes by the interaction of two nearest neighbour particles. We characterize the randomness in time by invoking the concept of WTD that measures the time between two consecutive kicks. This circuit model can be mapped to a classical system of many rotors with static kinetic energy and kicked nearest-neighbour interactions. As expected these kicks are not periodic but either random or quasi-periodic. We consider here two cases of WTDs, the case one assumes that the waiting time can become unbounded resulting in an infinitely long tail WTD profile. By contrast,  for case two, the waiting time can take some specific values leading to a sharp cut-off in WTD. For the  case one, we consider two random driving sequences, binomial and Poisson where the kinetic energy curves provide a good data collapse in the chaotic regimes by rescaling the time axis. This leads to a scaling relation for the diffusion constant as $D\sim K^2$, where $K$ is the kicking strength. For Poisson and binomial distributions, where the surprise $S$ depends on the distribution parameter, the diffusion constant is modified by the relation $D\approx SK^2$.
This modifies the diffusion constant for the periodic drive which is applicable for large $K$~\cite{sadia2022prethermalization}, since then the rotors move independently and one can neglect correlations. However, for the present aperiodic cases, the relation $D\sim K^2$ is valid even for smaller $K$ values. This behavior follows the fact that the breaking of time translation symmetry in a system generally opens many heating channels and destabilizes it rapidly~\cite{wen2021periodically}.

In addition, for both binomial and Poisson driving sequences, we calculate the lifetime of the so-called prethermal state numerically and both processes provide a relation $t^*\sim 1/K^2$ within numerical precision. For multi-polar drives with asymmetric interactions, the lifetime of the prethermal state is found as $t^*\sim 1/K^2$ in Ref.~\cite{yan2023prethermalization}. This result is supported by analytical arguments of the behaviour of the eigenvalues of the update matrix after linearization of the many-body kicked rotor Hamiltonian~\cite{yan2023prethermalization}. Interestingly, the lifetime of the prethermal state for our cases provides the same scaling relation as the lifetime they have found. Although we have considered generic random driving cases, unlike structured random multi-polar driving, we find a prethermal regime with the same prethermal lifetime scaling $t^*\sim1/K^2$ that proves the generality of the prethermal behaviour of a many-body rotor system. To be more precise, the algebraic relation of the $t^*$ and $K$ clearly suggests that the heating is not exponentially suppressed but rather algebraically suppressed. This distinct nature allows us to refer to this phase as pesudo-prethermal phase. With a close look at the collapsed curves of Figs.~\ref{fig:poisson_ke}, and \ref{fig:binomial_ke}, we observe that the heating time scales as $t^*\approx \tau/SK^2$. Further, using energy-time type uncertainty relation and numerical results of the lifetime of the pseudo-prethermal state, we find out the temperature of this prethemal state as $T^*\sim SK^2/\tau^2$. We have also made a connection between phase space entropy and surprise using time-dependent phase space probability density. This suggests to write the lifetime of the prethermal state as $t^* \sim \tau/\Delta \mb S_n ^2$, where $\Delta \mb S_n$ is the change in phase space entropy between two consecutive kicking events.

To study further, we consider the TM driving sequence where the waiting time is not unbounded rather than consider three values with equal probabilities i.e., a box distribution of WTD (see Fig.~\ref{fig:thue-morse_ke}(a)). As mentioned before, the heating time for this drive sequence scales as $t^*\sim e^{c/K}$, where $c$ is numerical constant and $K$ is the kicking strength. This is clear signature of the prethermal phase as noticed for Floquet case \cite{rajak2019characterizations}. However, in a previous work~\cite{yan2023prethermalization}, it has been reported that the lifetime of the prethermal state for TM drive has a scaling form $t^* \sim e^{\log(1/K)^2}$. This indicates that the lifetime and the heating time have different scaling forms for TM drive in a many-body rotor system.


\section{Conclusions}

In conclusion, we have studied how the prethermal behavior of a system of interacting classical rotors changes when 
it is driven either quasi-periodically or randomly. To accomplish our motivation, we consider two situations: 
(a) for case one, the driving sequences are purely random and the WTDs have tails prolonging to infinity, (b) for  case two, the driving sequence is quasi-periodic and the waiting times are finite with equal probabilities. For case one, we consider binomial and Poisson kicking sequences, whereas, for case two, we assume it to be TM sequence. We can actually tune the temporal randomness in the system by varying some parameters for random driving cases. We have also found analytical forms of WTD using the kicking sequences. Our findings show that case one (two) leads to an algebraic (exponential) suppression of heating in the pseudo- (regular) prethermal regime while diffusive regime also exhibits distinct behavior in terms of interaction parameters. Due to the randomness of drive, we introduce an energy-time type uncertainty relation that leads to writing an expression of the temperature of the so-called prethermal state combining the numerical results for random drive cases. We have also discussed the time-dependent phase space distribution and found a relation between phase space entropy and surprise of WTD. Finally, we find that the lifetime of the prethermal state varies as inverse to the square of change in phase space entropy between two consecutive kicking events.
Since our driving protocols include pure random driving cases, we can conjecture that the existence of the relatively short-lived prethermal-like regime is a generic phenomenon even for the aperiodic drives unlike the long-lived like the Floquet
prethermal regime ~\cite{rajak2018stability,rajak2019characterizations}.

\textit{Acknowledgments}---
AK thanks Aur\'elia  Chenu, Juan MR Parrondo and Carlos Mejia-Monasterio for discussions and comments. AK is funded by the Luxembourg National Research Fund (U-AGR-7239-00-C).

\appendix

\section{Probability Distribution}
\label{appendix_A}

Floquet driving is generalized to random driving by introducing WTD. The time between the kicks is always in integer multiples of the characteristic kick length $\tau$, as waiting time $\tau_w = l\tau$, where $l$ is an integer. We set $\tau=1$ for convenience. In this setting, $l$ plays the role of the waiting time. A periodic Floquet drive always has waiting time $l=1$ between two consecutive kicks. For a random driving, $l$ is interpreted as a random variable taken from a discrete distribution $P(l)$. We interpret the similarity between the distributions by the quantity called "surprise" of the distribution given as $S = -\sum_l P(l) \log P(l)$. This work considers two types of WTDs with unbounded and bounded profiles. For the case one, the waiting time can take any value between $0$ to $\infty$, and it goes up to only a finite cut-off value for case two. We consider $S\approx1$ for all the WTD so that we compare the findings from different WTD in a similar footing. 

\subsection{Unbounded WTD  with $l \in (1,\infty)$}

In this category, we consider discrete geometric and exponential distributions for the waiting time. The details of the distributions are given below.


\subsubsection{Discrete geometric distribution}

\begin{figure}[h]
    \centering
    \includegraphics[scale=0.25]{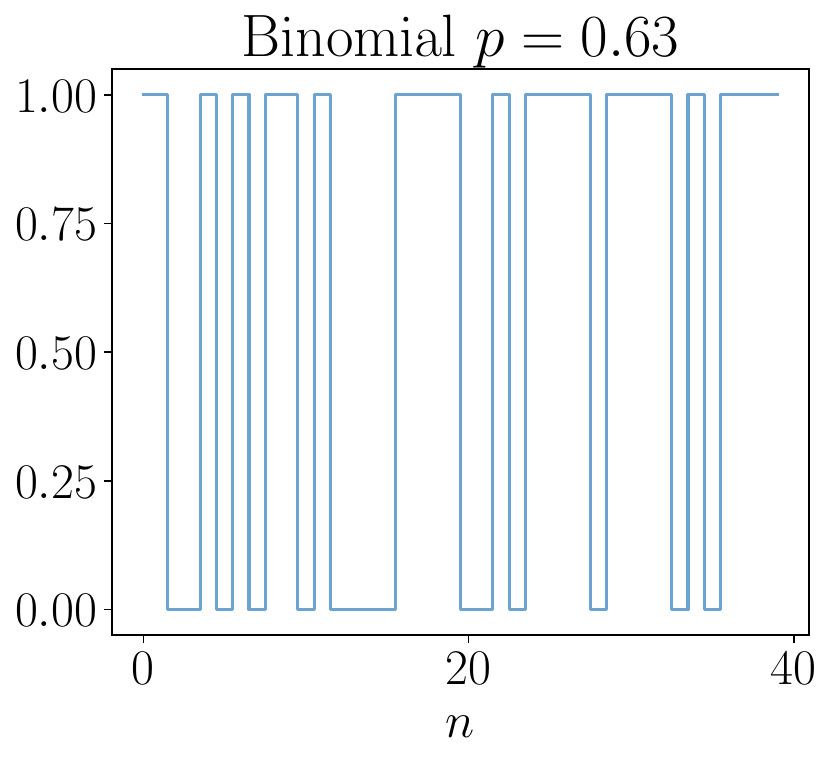}
    \includegraphics[scale=.25]{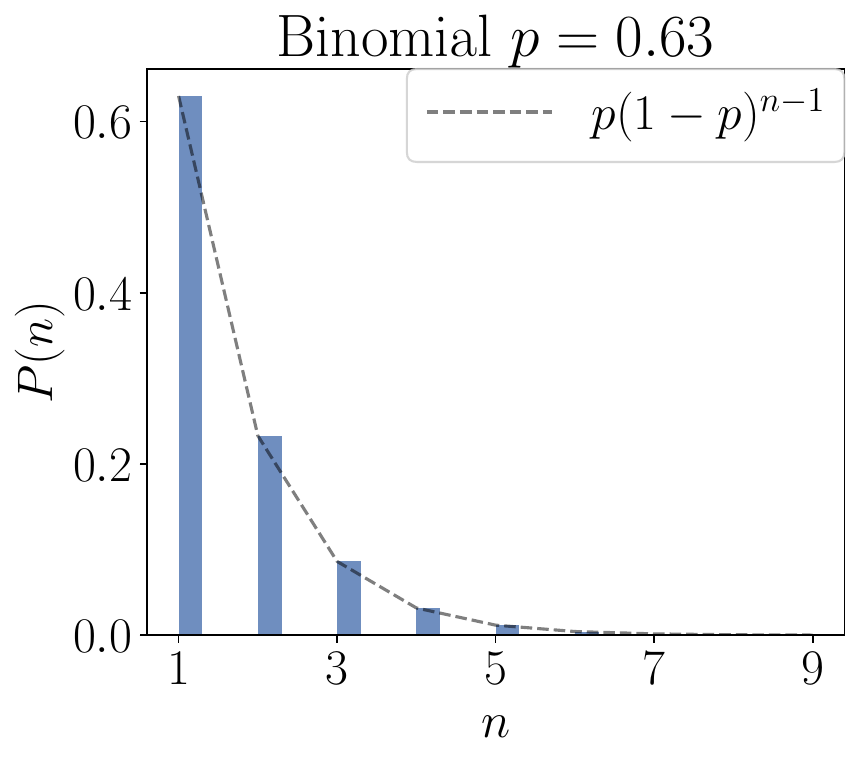}
    \caption{(a) Sequence from an Binomially distributed waiting time with $p=0.63$. (b) The corresponding distribution as in the text. The average surprise is $S=1.04$. }
    \label{fig:Binomial_0p63}
\end{figure}

\begin{figure}[h]
    \centering
    \includegraphics[scale=0.25]{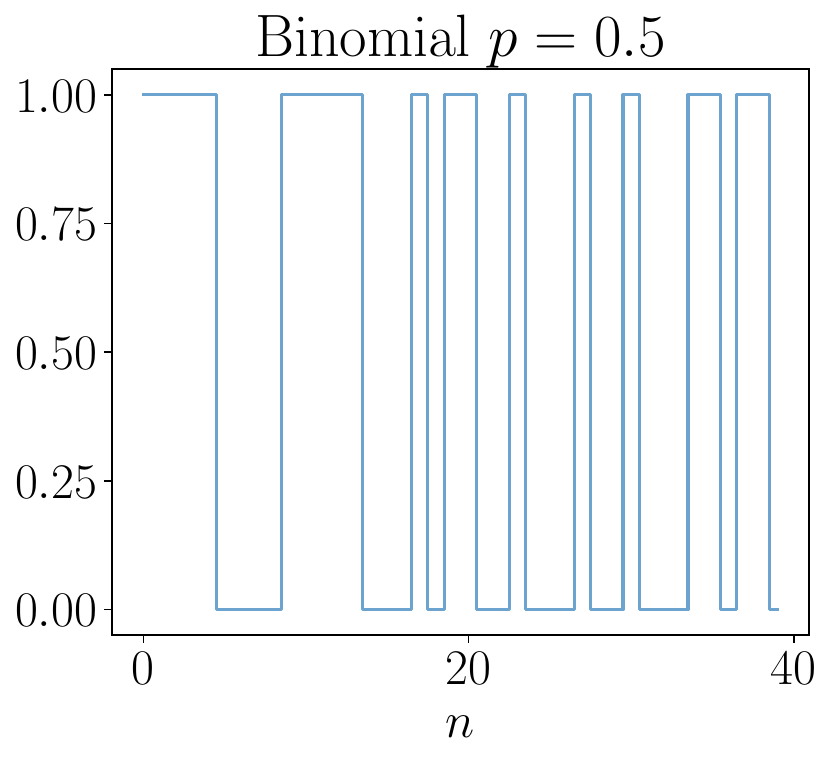}
    \includegraphics[scale=.25]{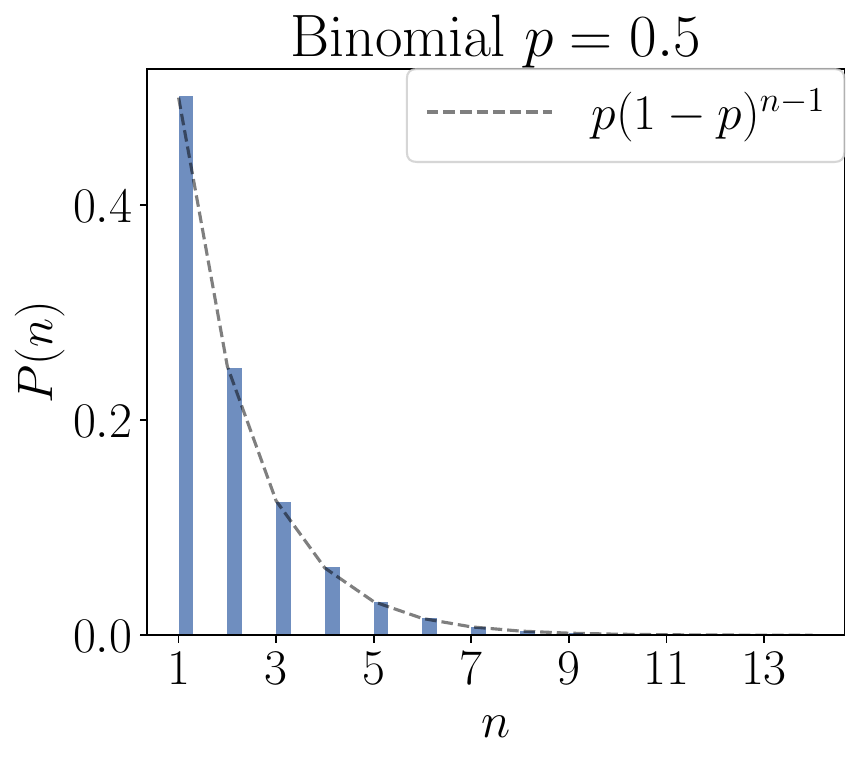}
    \caption{(a) Sequence from an Binomially distributed waiting time with $p=0.5$. (b) The corresponding distribution as in the text. The average surprise is $S=1.386$. }
    \label{fig:Binomial_0p5}
\end{figure}

In this case, the kicking sequence follows binomial probability distribution, and the corresponding WTD is given by $P(l)= p(1-p)^{l-1}$, where $l$ being the number of kicks. We have shown kicking sequences and the corresponding WTD for $p=0.63$ and $0.5$ in Figs.~\ref{fig:Binomial_0p63} (a,b) and \ref{fig:Binomial_0p5} (a,b), respectively. The value of probability of kicks $p=0.63$ leads to $S\approx1$ that we have followed for all the WTD.

\subsubsection{Discrete exponential distribution}

\begin{figure}[h]
    \centering
    \includegraphics[scale=0.25]{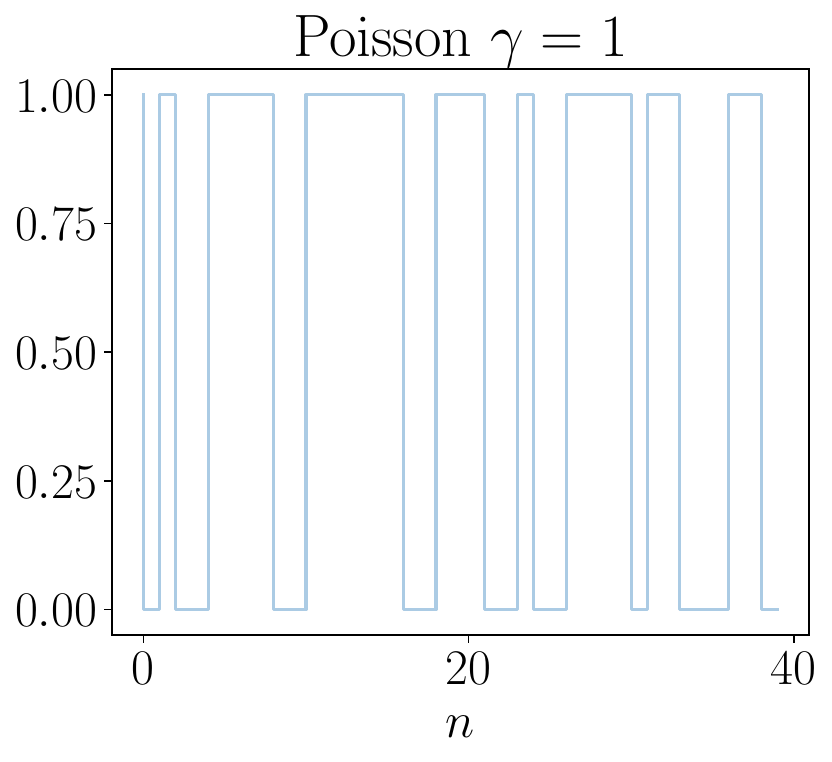}
    \includegraphics[scale=.25]{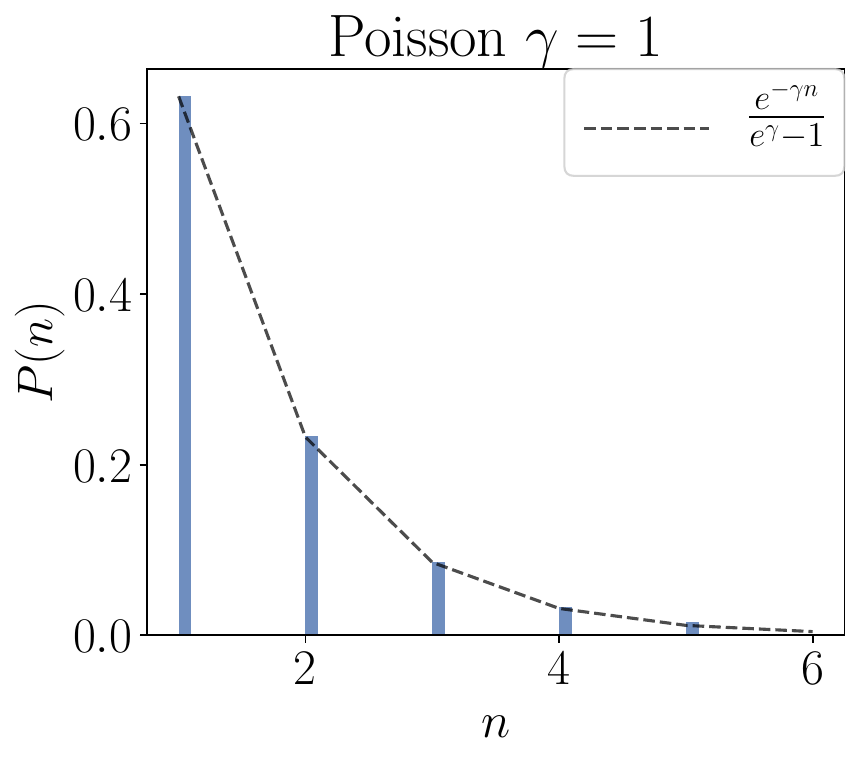}
    \caption{(a) Sequence from an exponentially distributed waiting time with $\gamma=1$. (b) The corresponding distribution as in the text. The average surprise is $S=1.04$. }
    \label{fig:PoissonSeqence}
\end{figure}

The standard Poisson process is a counting process which accounts for the number of kicks in a finite interval of time $[0,n]$ denoted by $N_n$ where the final time is $t= n$.
Let us say that the counts occur at times $T_1,T_2 \dots T_n$, then total number of $N_n = \sum_{i=1}^n \Theta(t-T_i) $, is $1$ when the argument is greater than zero or $0$ otherwise.
In a Poisson process, a further assumption is that the increments between the time steps have no memory, are independent of each other, and come from a stationary distribution. Here $N_n=k$ is a random variable which follows the distribution $$P(N_n = k) = e^{-\gamma n} \frac{(\gamma n )^k}{k!}  $$ described by the Poisson parameter $\gamma$. This distribution describes the probability of observing $k$ kicks in an interval of time $n$.
This provides the average number of kicks the system experienced after $n$ physical time as $\langle N_n\rangle = \gamma n$ with a variance equal to the mean. Then the distribution of waiting time between two kicks is given by, $$P(N_n = 0) = e^{-\gamma l},$$
where $l$ is the final integer time for which $N_n=0$. This leads to discrete exponential WTD $P(l)=Ce^{-\gamma l}$, where $C$ is the normalization constant. Considering the normalization condition $\Sigma_{l=1}^{\infty}  P(l)= 1$, $C$ is found to be $1/(e^{\gamma}-1)$. A sequence of kicks and the corresponding distribution are shown in Figs.~\ref{fig:PoissonSeqence} (a,b), respectively, for $\gamma=1$ so that the surprise is found to be $S\approx 1$.

 \subsection{Bounded WTD with $l \in (1, k$)}

In this case, the waiting time is bounded by a finite value. In this category, we consider, TM drive (see Sec.~\ref{thue-morse}) to examine the prethermal behavior of the system. The TM drive is the limiting case of the multipolar drive with $m\rightarrow\infty$. We discuss here different types of multipolar drives along with TM drive. 
Multipolar drives can be viewed as a sequence which is a random process (see Sec.~\ref{thue-morse} for details).


\subsubsection{Monopolar drive} 

\begin{figure}[h]
    \centering
    \includegraphics[scale=0.25]{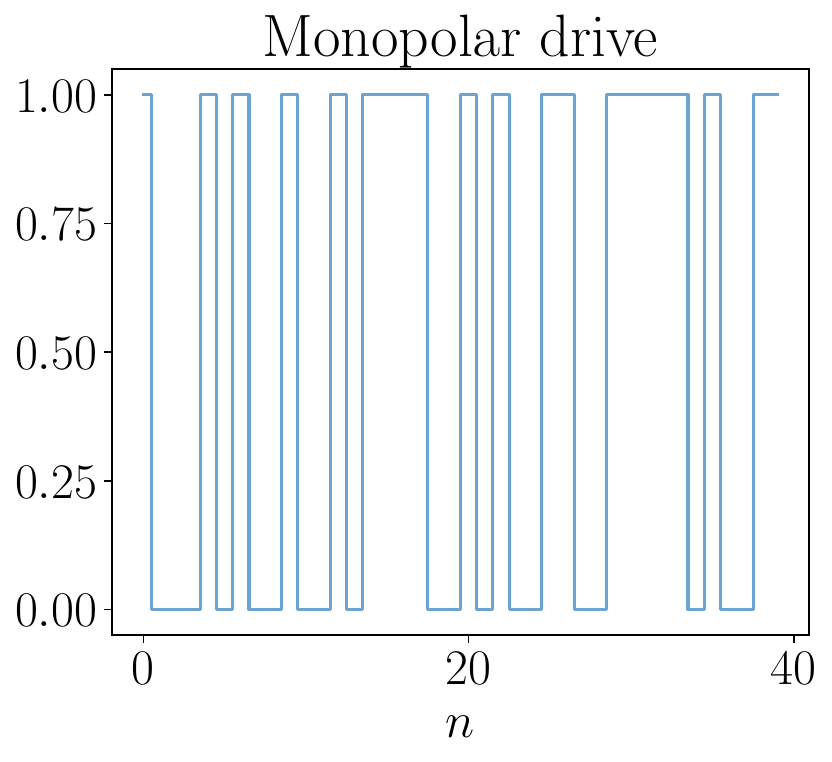}
    \includegraphics[scale=.25]{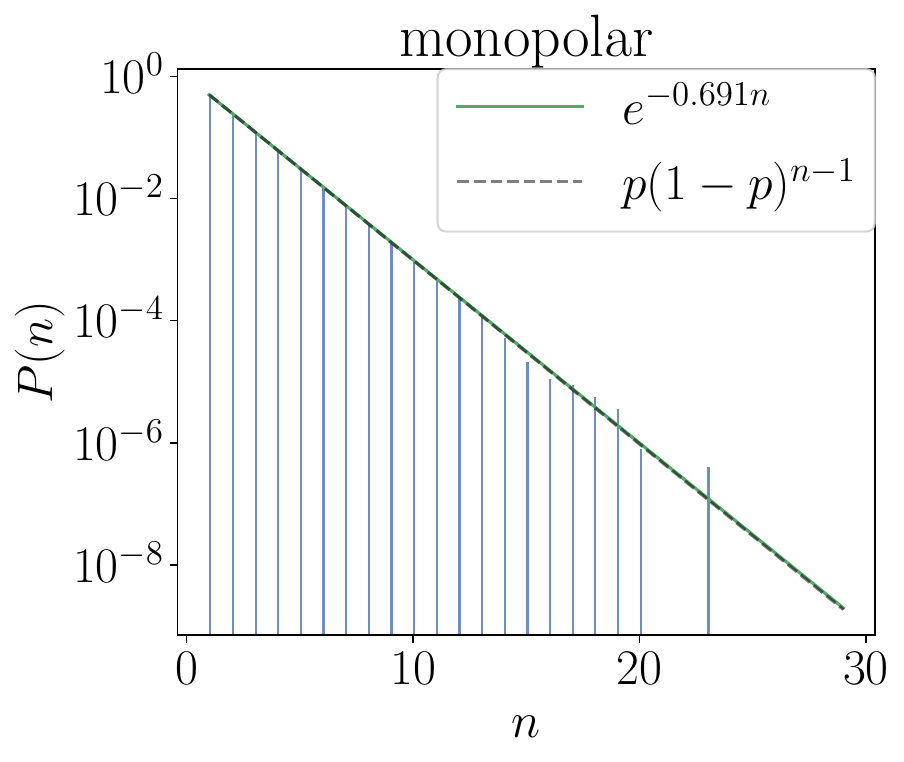}
    \caption{(a) Kicking sequence for a monopolar drive that is equivalent to binomial drive with $p=0.5$. (b) The distribution of the waiting time corresponds to the monopolar drive.}
    \label{fig:MonopolarSeqence}
\end{figure}

The monopolar drive as discussed in the main text is equivalent to selecting the kicks from a Binomial distribution. This corresponds to the waiting time taken from as that of the discrete Geometric distribution with $p=0.5$. The surprise for this case ones given by $S=1.386$ as shown in Fig.~(\ref{fig:Binomial_0p5}). The same amount of "surprise" is matched for an exponential waiting time with $\gamma \sim 0.691$. The kicking sequence and the corresponding WTD are shown in Fig.~\ref{fig:MonopolarSeqence} (a,b), respectively.

\subsubsection{Dipolar drive}
\begin{figure}[h]
    \centering
    \includegraphics[scale=0.25]{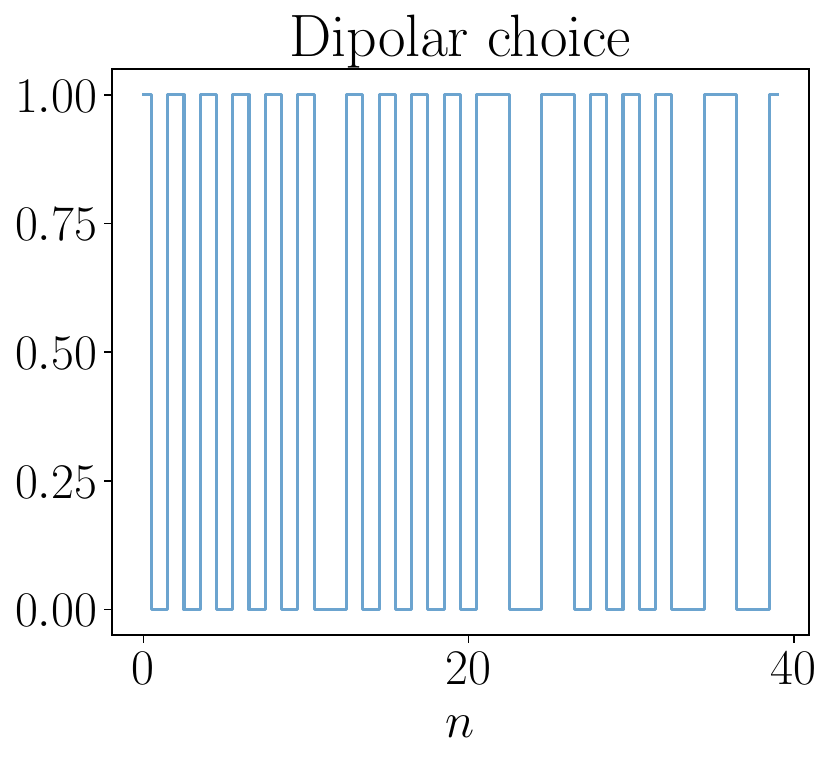}
    \includegraphics[scale=.25]{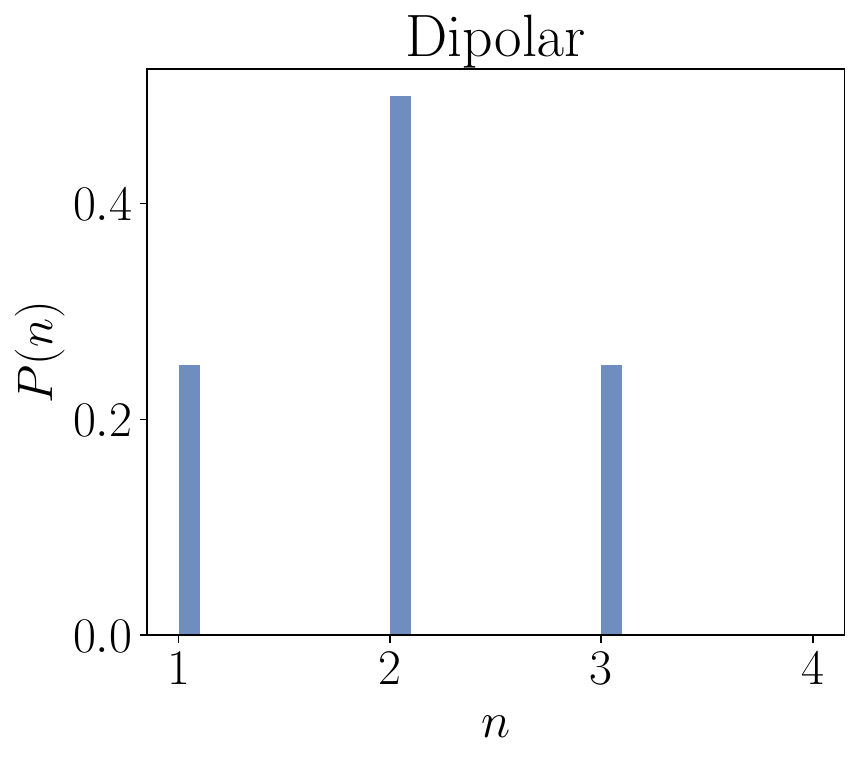}
    \caption{(a) A sequence of dipolar drive with (b) the corresponding WTD with a finite support. The surprise for this drive is found to be $S = 1.03972$. }
    \label{fig:DipolarSeqence}
\end{figure}

The dipolar drive as discussed in Sec.~\ref{thue-morse} is equivalent to selecting the kicks with discrete waiting time $l \in (1,3)$ with different probabilities. The WTD for this drive is given by
$P(l) =  \frac{1}{4} \delta (l-1) + \frac{1}{2} \delta (l-2)+ \frac{1}{4} \delta (l-3)$. The average and the variance of the distribution are found to be $\langle n \rangle = 2$ and $\langle n;n \rangle = 0.5$, respectively. The corresponding surprise is given by $S = 1.03972$. We have shown the kicking sequence and corresponding WTD in Figs.~\ref{fig:DipolarSeqence}(a,b), respectively.

\subsubsection{Thue-Morse drive}

\begin{figure}[h]
    \centering
    \includegraphics[scale=0.25]{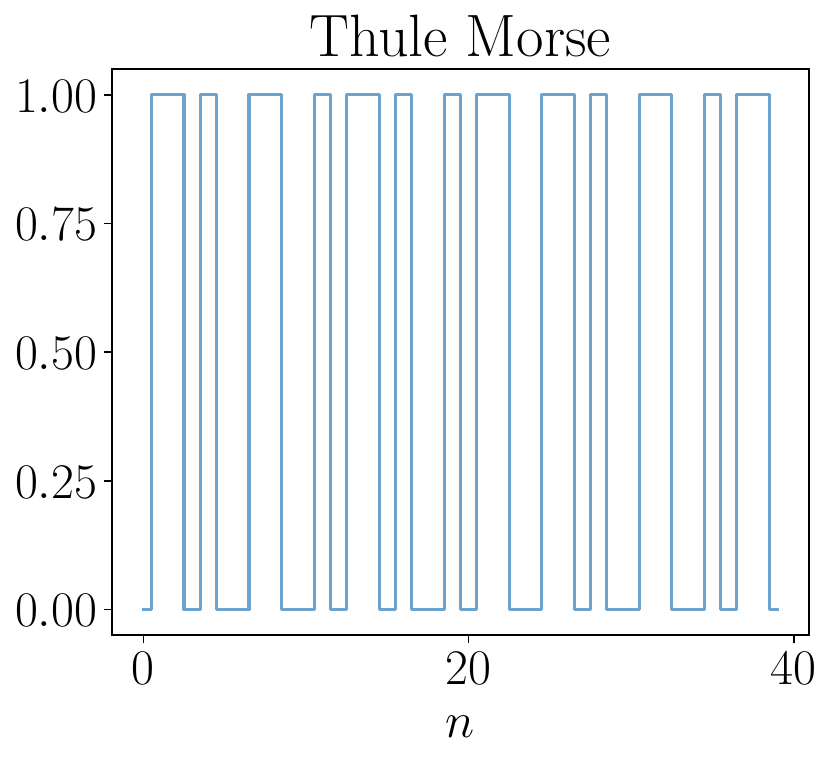}
    \includegraphics[scale=.25]{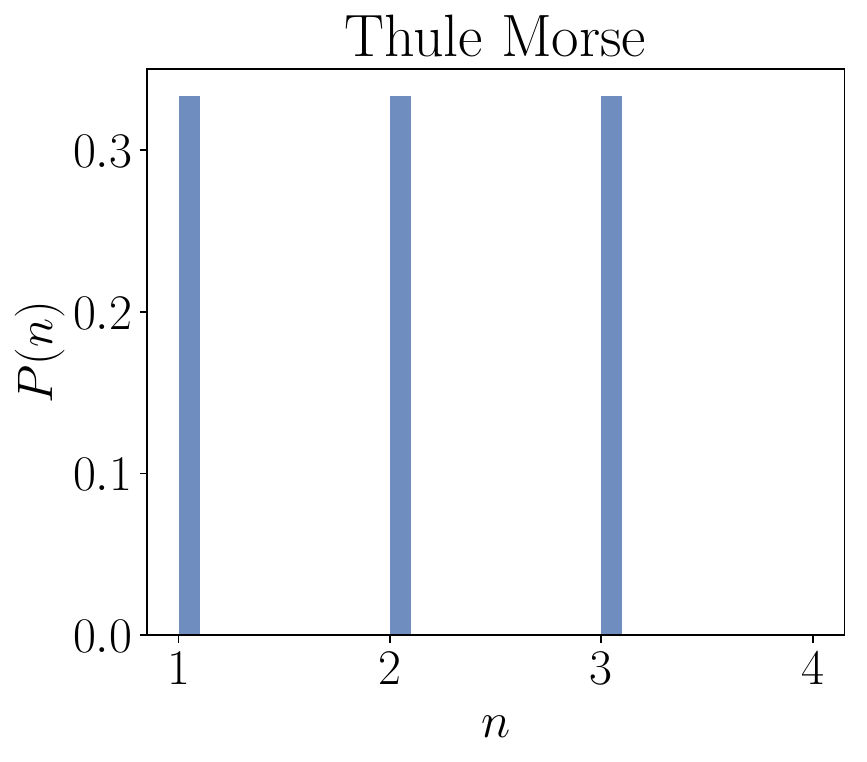}
    \caption{Thue-Morse driving sequence with WTD with a finite support. The surprise is $S = 1.098$. }
    \label{fig:ThuleMorseSeqence}
\end{figure}

This quasi-periodic drive is generated by Thue-Morse sequence. The waiting time is bounded by $l \in (1,3)$ and they are equally probable. This leads to the WTD $P(l) = \frac{1}{3} \sum_{i=1}^3 \delta (l-i)$ with the average $\langle l \rangle = 2$ and the variance $\langle l;l \rangle = 1$. The corresponding surprise is given by $S = 1.09861$. The driving sequence and corresponding distribution are shown in Fig.~\ref{fig:ThuleMorseSeqence} (a,b), respectively.

\section{Averaged phase slip dynamics}\label{app:phaseslip}

\begin{figure}
    \centering
    \includegraphics[scale=0.5]{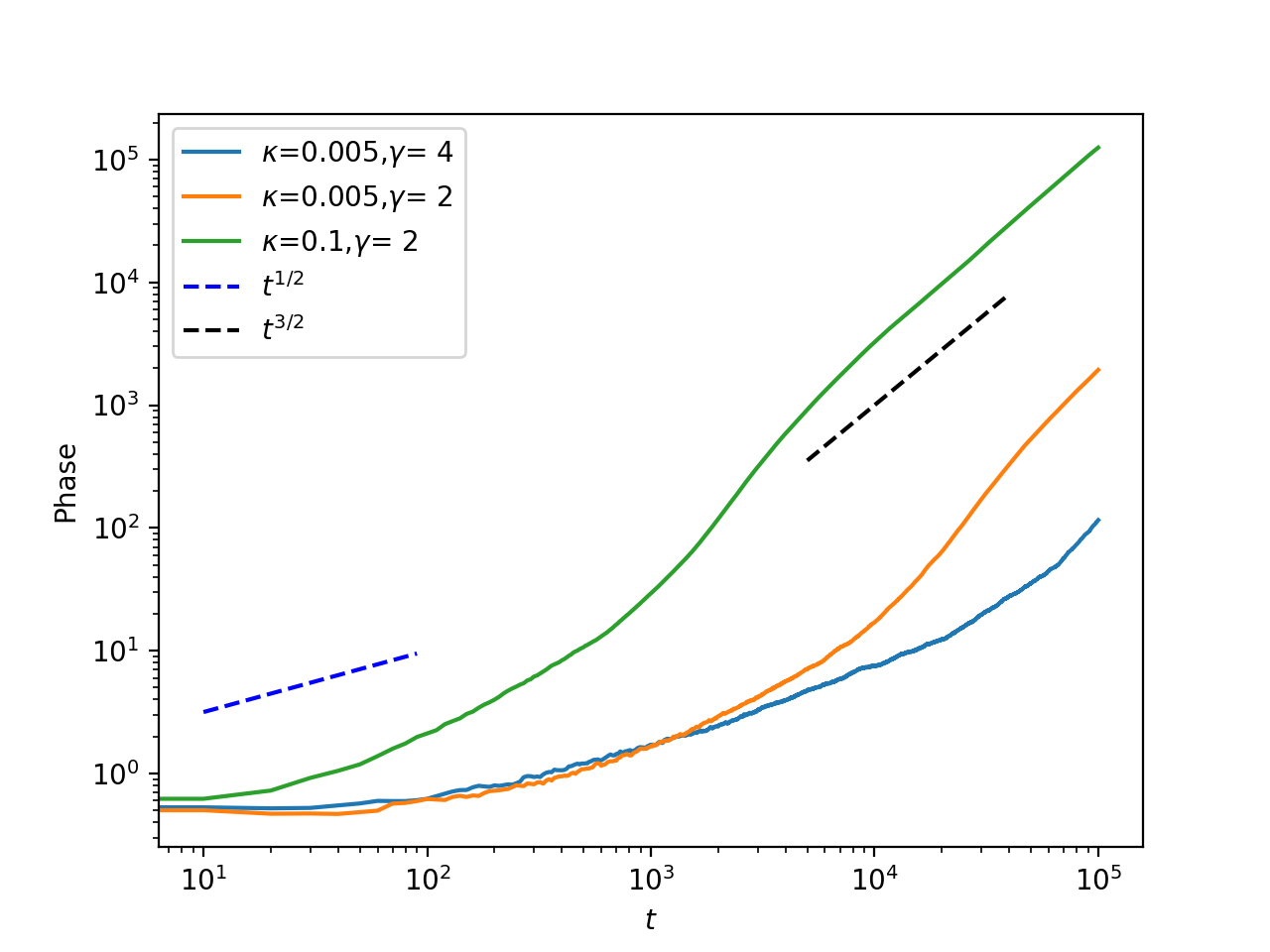}
    \caption{Averaged phase slips for the randomly driven system as a function of time by Poisson noise. The time evolution shows clear difference between inbetween the transition to the heating regime. For short times, the behaviour is $t^{1/2}$ while for heating the behaviour is $t^{3/2}$.}
    \label{fig:Phaseslip}
\end{figure}

The averaged phase slips count the average number of times the relative angle between adjacent rotors crosses $2\pi$. It is defined as 
$\avg{\phi} = \frac{1}{N} \sum_i {\rm Mod}[\phi_{i+1}-\phi_i,2\pi]$. It measures the dynamic transitions that occur in evolution as a proxy. In Fig.~\ref{fig:Phaseslip}, we see that the average number of phase slips changes from having a $t^{1/2}$ behaviour in the prethermal regime to a $t^{3/2}$ behaviour in the heating regime. Therefore we can differentiate the prethermal and heating regimes by the exponents of phase slip dynamics.


\section{Derivation evolution of averaged observables}\label{app:doublePoisson}

The discrete kick Hamiltonian over a time interval $\tau$ in Eq.~(\ref{mkr_eq}) can be viewed as a system driven by Poisson noise
\eqa{dH_\tau=\sum_i H_0(i) \tau -H_1(i) dN_\tau,} 
where $dN_\tau $ is a Poisson noise  such that rate $\avg{dN_\tau} = \gamma \tau$. $H_0$ denotes the non-interacting part while $H_1$ denotes the interacting part of the Hamiltonian. Now consider the mapping of Poisson noise to Gaussian noise in large $\gamma$ limit \cite{Wiseman1996},
\[\eta dt = (dN_t - \gamma d t)/|\sqrt{\gamma}|\].
The stochastic Hamiltonian
is given by \eqa{ {H} = {H}_0 + \gamma H_1 + \sqrt{2\gamma} \eta {H}_1}
where $\eta$ is a space time white noise with $\avg{\eta} =0$ and $\avg{\eta(t)\eta(t')} = \delta(t-t')$. We will move to the quantum picture for convenience and then use quantum-classical correspondence. Using the propagator $A_{t+dt} = U_{dt} A_t U^\dagger_{dt}$ and expanding using Ito-rules, the evolution of an averaged observable in the Heisenberg picture evolving with this Hamiltonian $H$ is given by the Lindblad master equations,
\eqa{d_t \avg{{A}} = i[{H}_0+\gamma H_1,\avg{A}] + 2\gamma [{H}_1,[{H}_1,\avg{A}]].}
The reverse passage from the more established Lindblad master equations to the stochastic Hamiltonian unravels the dissipative quantum dynamics. Replacing the commutators with Poisson brackets, $i[.,.] \to \{.,.\}$
we have for a classical observable $A$ evolving with classical stochastic Hamiltonian $H$.
 
\eqa{\partial_t \avg{A} = \{H_0 +\gamma H_1,\avg{A}\} -2\gamma \{H_1,\{H_1,\avg{A}\}\}}

The first term is the Ehrenfest equation of motion, reproducing the exact average equation of motion of the particle moving in the field.

\section{Generator for Poisson kicking}\label{app:poissongen}


Let the evolution of the phase-space density is given as 
$\partial_t \mb P(\Gamma_t) = -i\mc{L} \mb P(\Gamma_t)$ in the continuum, then
 $\mb P(\Gamma_t) = e^{-i\mc L t} \mb P(\Gamma_0)$. 
 Introducing a scaling variable $\lambda = t/\tau$,
 Remember $t = \sum_n \sigma_n$, is a random variable. For a given $t$, the number of kicks $n$ is random with a distribution $p(\alpha,t)$. The average density function at time $t$ is then a convolution of the phase space distribution with the random distribution, 
 \eqa{\mb P(t) = \int_0^\infty dt \mb P_s(\Gamma_t) p(\alpha,t),}
Introducing scaled variable $P(\alpha,\lambda=t/\tau) = \tau p(\alpha,t)$, we have 
  \eqa{\mb P(t) = \int_0^\infty d\lambda \mb P_s(\Gamma_\lambda) P(\alpha,\lambda).}
The propagator is then given as,
\eqa{G_\alpha(\tau) = \int_0^\infty d\lambda P(\alpha,\lambda) e^{-i\mc L \lambda \tau}.}
Let us define the scaled Poisson probability distribution for collisions as:
\eqa{P(\alpha,\lambda) = \frac{e^{-\lambda} \lambda^{({\alpha}-1)}}{\Gamma(\alpha)},}
gives,
\eqa{G_\alpha(\tau) = \int_0^\infty d\lambda \frac{e^{-\lambda(1+i \mc L \tau)} \lambda^{({\alpha}-1)}}{\Gamma(\alpha)} = (1+i \mc L \tau)^{-\alpha}.}
For $\alpha \to 0$, we recover the continuous evolution.
The evolution of the phase space is then $\mb P(\Gamma_{t-\tau}) - \mb P (\Gamma_t) = -i \mc L \mb P(\Gamma_t) \tau$. This shows for Poisson WTDs, the original propagator becomes exact.

\bibliography{ref}

\end{document}